\documentclass[aps,twocolumn,superscriptaddress]{revtex4-1}
\usepackage{graphics} 
\usepackage{epsfig} 
\usepackage{amsmath}
\usepackage{color}

\begin{document}
\title{Collective excitations of rotating dusty plasma under quasilocalized charge approximation of strongly coupled systems}
\author{Prince Kumar}
\email{kumarprincephysics@gmail.com}
\affiliation{Institute for Plasma Research, HBNI, Bhat, Gandhinagar, India, 382428}
\affiliation{Homi Bhabha National Institute, Training School Complex, Anushakti Nagar, Mumbai 400085, India}
\author{Devendra Sharma}
\affiliation{Institute for Plasma Research, HBNI, Bhat, Gandhinagar, India, 382428}
\affiliation{Homi Bhabha National Institute, Training School Complex, Anushakti Nagar, Mumbai 400085, India}
\date{\today}

\begin{abstract}
Collective excitations of rotating dusty plasma are analyzed 
under the quasi localized charge approximation (QLCA) framework for strongly 
coupled systems by explicitly accounting for the dust rotation in the analysis. 
Considering the firm analogy of magnetoplasmons with ``rotoplasmons'' 
established by the recent rotating dusty plasma experiments, the relaxation 
introduced by the rotation in their strong coupling and 2-dimensional (often 
introduced by gravitational sedimentation) characteristics is emphasized in 
their dispersion. 
Finite rotation version of both strong and weak coupling dispersions is 
derived and analyzed, showing correspondence between a `faster rotating but 
weakly coupled' branch and its strongly coupled counterpart, 
relevant to both magnetized and unmagnetized dust experiments, in gravity or
microgravity conditions. The first correspondence between their measurements 
in rotating plasmas and the QLCA produced dispersions in a rotating 
frame, with an independent numerical validation, is presented in detail.
\end{abstract}



\maketitle
\section{Introduction \label{introduction}}
Collective excitations in systems with strongly interacting charged 
particle constituents are fundamental to understanding the evolution of 
many physical systems, including planetary systems
\citep{yaroshenko2007,Yaroshenko,Shukla_2003}, accretion disks
\citep{nekrasov2009}, binary stars and ultra dense crust of the neutron stars \citep{chabrier2002dense}. 
Strongly coupled phase of matter efficiently models physical state of not only 
astrophysical systems such as ion fluid in white dwarfs \citep{koester1986evolution}, neutron 
star exteriors but also many laboratory scale systems that are dynamically 
close to them, such as localized ions\citep{willitsch2008chemical} in extended cryogenic traps or heavily 
charged dust grains \citep{pramanik2002experimental,quinn2000experimental} suspended in electron-ion plasmas. 

While the rotation is an inseparable ingredient of dynamics of most of 
the astrophysical systems, interplay of rotation with extremely large magnetic 
field  present in them provides vital clues about their 
dynamical state as well as their internal structure, for example pulsating 
radiations from pulsars because of misaligned magnetic and rotational axes \citep{ghosh2007rotation}
that led to their discovery \cite{hewish1968observation}.
Achieving the magnetization of heavier species in strongly coupled system 
realized in normal laboratory conditions remains a considerable challenge.
For example, the dusty plasma experiments where observable dust magnetization 
needs considerably high magnetic field \cite{uchida2004}, hence dedicated 
facilities to achieve this goal \cite{thomas2015magnetized,thomas2012magnetized}. 
A remarkable dynamical analogy between the effects of rotation and 
magnetization in such systems although exists and is illustrated in some recent
laboratory dusty plasmas experiments where an ``effective magnetization'' 
resulted from an induced dust rotation \cite{kahlert2012}. 
As estimated, for example, by K\"{a}hlert et al. \citep{kahlert2012}, a 
rotation frequency of about 10 Hz in a typical dusty plasma (dust charge
$\sim 10^4e$ and mass$\sim 10^{-12}$ kg) generates effects equivalent to 
magnetic field exceeding $10^{4}$ T.
Advanced experiments 
with dust rotation have also demonstrated excitation of dust magnetoplasmons 
in such ``effectively magnetized'' dust components \cite{hartmann2013}.
Addressing the analytical approach to these results, by explicitly 
admitting dust rotation in one of the standard dynamical models of strongly 
coupled systems, and their quantitative comparison with the experimentally 
observed dispersion is the subject of the present paper.
The analysis also applies to perturbations of general equilibrium dust vortex 
solutions in unmagnetized \cite{manjit} as well as weakly magnetized plasma 
conditions \cite{prince}.

The collective excitations in strongly to weakly coupled systems are widely 
formulated and studied under the quasi-localized-charge approximation (QLCA)
approach \cite{Golden,kalman1990response,golden1992dielectric}. The QLCA formulation appropriately reduces into usual 
fluid formulation under a random phase approximation (RPA) regime 
\cite{Golden,hou2009wave}. 
Strong coupling effects in presence of a magnetic field have also been 
addressed under the QLCA formulation \citep{kalman1990response,jiang2007theoretical}, 
indicating that its results may be applicable to the ``effectively magnetized''
dusty plasmas.

Essential to the characterization of collective excitations of effective 
magnetoplasmons is to ascertain their dispersion, as observed in the 
experiments, against the analytic dispersion relation produced by the 
theory of strongly coupled rotating dust system.
An existing QLCA approach that considered dust interacting by Yukawa 
interaction was done by Jiang et al.\citep{jiang2007theoretical}  who also considered magnetic 
field and therefore is relevant to the cases with dust rotation.
The result from the Jiang et al. however showed that the agreement of 
magnetized QLCA results with computer simulations is achievable 
for strongly coupled limit solutions. Specifically, the saturation displayed 
in the frequency of the dispersive excitation at larger wave numbers 
remained achievable only for strong coupling and is absent for 2D RPA limit.
In somewhat contrast, the saturation is measured with relatively weak 
coupling (liquid phase) regime in the recent ``effectively magnetized'' 
rotating dusty plasma experiment \cite{hartmann2013} and also in the 
corresponding computer simulations, both of which were interpreted based on 
existing QLCA approaches. 

Of the immediate interest therefore is to workout a QCLA formulation which 
explicitly accounts for dust rotation in a Yukawa system, and also derive
its limiting cases, for example, that of the weakly coupled (or RPA) limit 
where it yields equivalence to the regular explicitly rotating ``dust fluid'' 
formulation, which is done in this paper. 
Not limiting our analysis to characterizing the analytically obtained
dispersion relations, we also compute the multi-component rotating dust 
fluid model solutions, producing their independent agreement with analytical 
dispersions, thus simultaneously validating the accuracy of the analytical 
results.

The analytical and numerical analysis of the rotating plasma dispersion 
presented here shows that if the weakly coupled limit is considered 
with distant axial boundaries (3D effects), the frequency saturation not 
only becomes achievable in the rotating dusty plasma but the saturated 
frequency values also show quantitative agreement with those recovered 
in the recent experiments of the rotating dusty plasmas \cite{hartmann2013}. 
This behavior indicates that a dust medium with large enough coupling 
parameter $\Gamma$, when 
driven to a nonequilibrium state by an external rotation, not only displays 
weak coupling like effects but also a relaxation from the boundary effects 
such that the plasma bulk strongly limited by axial boundaries still displays 
nearly 3D characteristics. The rotation of the dust thus additionally 
appears to relax the 2D effects (2D sedimentation, usually generated by 
the gravity\citep{khrapak2016complex}) on the dust cloud. 
It is therefore concluded that this nonequilibrium rotating system not only 
shows isomorphism with magnetized dust regime but perhaps also with its
microgravity regime, rendering the latter realizable also in the regular 
ground based laboratory experiments, if dust is subjected to a 
rotation. 
The analysis therefore motivate studies both in magnetized as well as in 
microgravity conditions in order to examine their similarity with 
ground based rotating dusty plasma experiments. The analysis may thus 
serve as a link 
between ground based magnetized dusty plasma experiments, like MDPX, and 
the International Space Station (ISS) based experiments in microgravity 
conditions \citep{pustylnik2016plasmakristall,fortov2003dynamics,khrapak2016complex}.

The present paper is organized as follows. The QLCA formulation, explicitly
accounting dust rotation, is developed in the Sec.~\ref{qlca}.
The first characterization of dispersion relation for a strongly coupled
rotating dusty plasma with Yukawa interaction is also done in Sec.\ref{qlca}
showing a tendency in RPA limit rotating dust dispersion to 
increasingly agree with the strongly coupled dispersion at stronger rotation.
The RPA limit dispersions are shown recoverable from a multi-component model 
with rotating dust fluid in Sec.~\ref{qlca}A. This model is solved 
numerically to validate the characterization of the RPA limit dispersion 
relation which is done using the set of parameter derived from recent 
rotating dusty plasma experiment in Sec.~\ref{solutions}B. A general 
characterization of the dispersion, independent of the experimental parameter 
sets, is presented in Sec.~\ref{solutions}C with respect to variation in 
rotation frequency and screening by the background plasma. Finally, the issue 
of strongly coupled rotating dust in experiments showing better agreement 
with RPA limit dispersion with increasing rotation frequency is addressed 
qualitatively in Sec.\ref{solutions}D.
%
\section{The quasi-localized-charge approximation (QLCA) for rotating 
dusty plasma\label{qlca}}
\begin{figure}[hbt]
 \centering
 \includegraphics{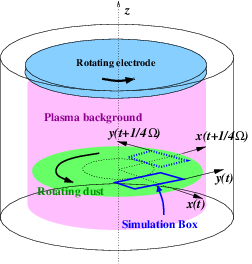}
	\caption{Schematic of the rotating dusty plasma setup. 
	The rectangular simulation zone located in the frame co-rotating 
	with the dust, is drawn at time $t$ (solid line) 
	and at quarter of the rotation period after time $t$ (dotted line) 
	in the laboratory frame.}
\label{schematic}
\end{figure}

The QLCA (quasi-localized-charge approximation) is very general microscopic 
model for the description of linear dielectric response and collective mode 
dispersion in strongly coupled systems. It assumes that charge particles, in 
strongly coupled system, are quasi-localized around the equilibrium position 
in corresponding potential wells and respond linearly to small 
external perturbing fields. 
Our approach to analyze the spectral properties 
is aimed to address the laboratory experiments having rotating dust component 
\cite{hartmann2013}. The strongly coupled dusty plasma in QLCA approach is 
well represented by a one component plasma with strong particle charge, 
namely, the dust, in a Yukawa potential produced by the shielding 
effect of the background electron and ion species.
As in the non-rotating QLCA treatment 
\cite{Golden}, 
we begin by writing the microscopic equation of motion for the single 
particle, in the absence of external force and collisions, however in 
the frame rotating with the dust cloud. We therefore consider a local 
coordinate system 
in the dust cloud such that the axis of rotation is aligned to $\hat{z}$.
We write our microscopic equation of motion of the dust 
particle located at the position $r_{i}$, 
for the component $r_{i\mu}$ aligned to the direction $\mu~(=x,y)$, 
\begin{eqnarray}
\begin{split}
%
%
%
%
%
	m_{\rm d}\frac{\partial^{2}{r}_{i\mu}}{\partial t^2}=
	\sum_{j} {K}_{ij\mu\nu}r_{j\nu}
	-2m_{\rm d}\left[{\bf \Omega} \times \frac{\partial {\bf r}_{i}}{\partial t}\right]_{\mu}\\
	-m_{\rm d}[{\bf \Omega}\times({\bf \Omega}\times {\bf r}_{i})]_{\mu} 
	-\frac{\partial V}{\partial r_{\mu}}=0,
\end{split}
	\label{particle-eq}
\end{eqnarray}   
where second and third terms in right hand side are Coriolis force and 
centrifugal force, respectively. The quantity $V$ is the confinement potential 
whose gradient along $\mu$ balances the corresponding component of the 
centrifugal force in the typical equilibrium condition \cite{nekrasov2009}. 
The dynamical tensor $K_{ij\mu\nu}$, with dimension index 
$\mu,\nu=x,y,z$ (index $i,j$ only enumerate particles), represents the 
effect of the inter-particle interaction in 
a Yukawa system which, in the non-retarded limit (velocity of electromagnetic 
waves $c\rightarrow \infty$)
is given by \cite{kalman1990response},
\begin{eqnarray}
%
%
%
	K_{ij\mu\nu}=(1-\delta_{ij})
	\frac{\partial^{2}}{\partial r_{i\mu}\partial r_{j\nu}}
	\phi(|{\bf r}_{i}-{\bf r}_{j}|)\\
	-\delta_{ij}\left\{\sum_{l=1}^{N}(1-\delta_{il})
	\frac{\partial^{2}}{\partial r_{i\mu}\partial r_{l\nu}}
	\phi(|{\bf r}_{i}-{\bf r}_{l}|)\right.\\
	\left.
	+\int d^{2}r'
	\frac{\partial^{2}}{\partial r_{i\mu}\partial {r'_{\nu}}}
	\frac{\rho_{\rm b}({\bf r'})}{\rho_{0}}
	\phi(|{\bf r}_{i}-{\bf r'}|)
	\right\},
\end{eqnarray}   
where $\rho_{\rm b}({\bf r'})$ is the charge density of the background plasma, $\rho_{0}$ is unperturbed dust charge density and potential $\phi$ has the form,
\begin{eqnarray}
	\phi(|{\bf r}_i-{\bf r}_j|)=e^{-\kappa_{b}|{\bf r}_{i}-{\bf r}_{j}|}\frac{Z^{2}e^{2}}{|{\bf r}_{i}-{\bf r}_{j}|},
\end{eqnarray}   
with  
\begin{eqnarray}
	\kappa_{b}^{2}=\sum_{A}4\pi Z_{A}^{2}e^{2}n_{A}\beta_{A}.
\end{eqnarray}   
The subscript $A$ corresponds the species in the background plasma, 
specifically electron and ions in a typical dusty plasma.
In its Fourier transformed form, the potential $\phi({\bf r-r'})$ therefore 
reads,
\begin{eqnarray}
	\phi(k)=\frac{4\pi Z^{2}e^{2}}{k^{2}+\kappa_{b}^{2}}.
\end{eqnarray}   
We now consider a small perturbation $\xi_{i\mu}$ of the equilibrium 
location $r_{i\mu}$ 
in order to explore the linear wave-like response of the system. 
Accordingly, the Eq.~(\ref{particle-eq}) in the 
frequency domain, obtained by the Fourier transformation is, 
\begin{eqnarray}
\begin{split}
%
%
	-m_{\rm d}\omega^{2}\xi_{i\mu}(\omega)-\sum_{j}K_{ij\mu\nu}\xi_{j\nu}(\omega)
	-i\omega 2m_{\rm d}~\epsilon_{\nu\delta\mu}{\Omega_{\nu}}{\xi}_{i\delta}(\omega)\\
	=ZeE_{\mu}(\xi_{i\mu},\omega).
	\label{particle-eq1}
\end{split}
\end{eqnarray}   
Note that $K_{ij\mu\nu}$ is independent of $\omega$ in the non-retarded limit 
and $\epsilon_{\nu\delta\mu}{\Omega_{\nu}}{\xi}_{i\delta}(\omega)$ represents 
the cross product ${\bf \Omega} \times {\bf \xi}_{i}(\omega)$ in the index 
notation, with $\epsilon_{\nu\delta\mu}$ having its conventional values with 
respect to the order $\mu\nu\delta$ of the indices, each taking values 
$x,y$ and $z$, respectively. 
Summation is implied over the repeated indices.
Considering that the 2D problem is solved in the plane perpendicular 
to ${\bf \Omega}$ (aligned to $\hat{z}$) and corotating with the dust, 
as described by schematic 
Fig.~\ref{schematic}, the index $\mu$ takes only two values,
$x$ and $y$, hence the cross product contributes only one term 
containing $\delta\neq\nu,\mu$. 
%

Considering their spatially extended distribution, displacements 
$\xi_i(\omega)$ can now be Fourier expanded in the series of 
vectors ${\bf q}$ as, 
\begin{eqnarray}
\begin{split}
	\xi_{i}(\omega)=\frac{1}{\sqrt{Nm_{d}}}\sum_{\bf q}\xi_{\bf q}(\omega)e^{i{\bf q}\cdot {\bf r}_{i}}.
	\label{expansion-xi}
\end{split}
\end{eqnarray}   
Similarly, expression of $K_{ij\mu\nu}$ in these terms 
\cite{kalman1990response} would be,
\begin{eqnarray}\nonumber
%
%
	K_{ij\mu\nu}=\frac{1}{V_{dD}}\sum_{\bf q} 
	q_{\mu}q_{\nu}\psi_{dD}({\bf q},\omega)
	[e^{i{\bf q}\cdot({\bf r}_i-{\bf r}_j)}\\
	-\delta_{ij}e^{i{\bf q}\cdot{\bf r}_i}n_{q}+
	\delta_{ij}\frac{n_{b{\bf q}}Z_b}{Z}e^{i{\bf q}\cdot{\bf r_{i}}}].
	\label{expansion-K}
\end{eqnarray}   
%
Using the dusty plasma quasi-neutrality condition we get $n_{b}Z_{b}=NZ$, 
where $Z_b$ is average charge per part in the background continuum if it 
is divided into same number of parts as the number of dust particles $N$.
Additionally, if the background plasma has no spatial structures, the factor
$e^{i{\bf q}\cdot{\bf r_{i}}}$ reduces to $\delta_{\bf q}$ and $n_{b{\bf q}}$ 
becomes independent of ${\bf q}$. Using these two conditions, the definition 
(\ref{expansion-K}) reduces to,
\begin{eqnarray}\nonumber
%
%
	K_{ij\mu\nu}=\frac{1}{V_{dD}}\sum_{\bf q} 
	q_{\mu}q_{\nu}\psi_{dD}({\bf q},\omega)
	[e^{i{\bf q}\cdot({\bf r}_i-{\bf r}_j)}\\
	-\delta_{ij}e^{i{\bf q}\cdot{\bf r}_i}n_{q}+
	N\delta_{ij}\delta_{\bf q}],
	\label{expansion-K1}
\end{eqnarray}   
where the subscript $dD$ in $V$ and $\psi$ defines the dimensionality of 
the system by takes values $dD=2D,3D$.
The general definition of the potential $\psi_{dD}$ is \cite{Golden},
\begin{eqnarray}
	\psi_{dD}(q,\omega)=\left[1-\left(\frac{\omega^2}{q^2c^2}\right)\right]^{(3-d)/2}\phi_{dD}(\omega),
\end{eqnarray}   
such that in 3D, $\psi_{3D}=\phi_{3D}$.

In order to examine an excitation with vector ${\bf k}$ in the dust cloud,
we follow the standard QLCA prescription \cite{Golden} and substitute 
(\ref{expansion-xi}), (\ref{expansion-K}), 
into Eq.~(\ref{particle-eq1}), multiply it with 
$\exp{(-i{\bf k}\cdot {\bf r}_{i})}$ 
(Note that $n_{\bf k}=\sum_{i}\exp{(-i{\bf k}\cdot {\bf r}_{i})}$) 
and sum over $i$, for $N$ particles. 
This sequence of operations produces,
\begin{eqnarray}
\begin{split}
	-m_{\rm d}\omega^2
	\sum_{\bf q}\xi_{{\bf q}\mu}(\omega)
	n_{k-q}\\
	-\frac{1}{V_{dD}}\sum_{\bf q,p} 
	q_{\mu}q_{\nu}\psi_{dD}({\bf q},\omega)
	[n_{k-q} 
	n_{q-p}\\
	-n_{k-q-p}n_{q}
	+N\delta_{q}n_{k-p}]
	\xi_{\bf p\nu}(\omega)
	\\
	+2i\omega m_{\rm d}~\epsilon_{\nu\delta\mu} \Omega_{\nu}
	\sum_{\bf q}\xi_{\bf q\delta}(\omega)
	n_{k-q}\\
	=Zen \sqrt{\frac{m_{d}}{N}}\sum_{\bf q}E_{\bf q}(\xi_{\bf q\mu},\omega)n_{k-q}.
 \end{split}
\end{eqnarray}   
We now apply the 
central
assumption of QLCA formulation and replace the quantities 
$n_{k-q}, n_{k-q}n_{p-q}$ and  $n_{k-q-p}n_{q}$ by their ensemble averages,
\begin{eqnarray}
	\langle{n_{\bf k-q}}\rangle&=&N\delta_{\bf k-q},\\
	\langle{n_{\bf k-q}n_{\bf q-p}}\rangle&=&NS(|{\bf k-q}|)\delta_{\bf k-p}
	+N^{2}\delta_{\bf k-q}\delta_{\bf q-p},\\
	\langle{n_{\bf k-p-q}n_{\bf q}}\rangle&=&NS(|{\bf q}|)\delta_{\bf k-p}
	+N^{2}\delta_{\bf q}\delta_{\bf k-p},
\end{eqnarray}   
where $S({\bf q})$ is static structure function.
In the last step, we substitute these ensemble averages into the respective 
terms and sum over indices, wherever possible,
As a result the equation further becomes,
\begin{eqnarray}
\begin{split}
	-m_{\rm d}\omega^2
	\xi_{{\bf k}\mu}(\omega)\\
	-\frac{N}{V_{dD}}
	k_{\mu}k_{\nu}\psi_{dD}({\bf k},\omega)
	\xi_{\bf k\nu}(\omega)\\
	-\frac{1}{V_{dD}}\sum_{\bf q} 
	q_{\mu}q_{\nu}\psi_{dD}({\bf q},\omega)\\
	\times [S(|{\bf k-q}|)-S(|{\bf q}|)]
	\xi_{\bf k\nu}(\omega)\\
%
%
%
%
	+2i\omega m_{\rm d}~\epsilon_{\nu\delta\mu} \Omega_{\nu}
	\xi_{\bf k\delta}(\omega)\\
	=Zen\sqrt{Nm_{d}}  E_{\bf k}(\xi_{\bf k\mu},\omega).
\end{split}
\end{eqnarray}   

This, in absence of external electric field $E$, yields a macroscopic equation 
for $\xi_{{\bf k}\nu}$ 
of the form, 
\begin{eqnarray}
	\nonumber
	[\omega^{2}\delta_{\mu\nu}-C_{\mu\nu}({\bf k},\omega)]
	\xi_{{\bf k}\nu}(\omega)
	+2i\omega~\epsilon_{\nu\delta\mu} \Omega_{\nu} \xi_{\bf k\delta}(\omega)\\
	=0.
\end{eqnarray}   
Which can be written in a more general form as,
\begin{eqnarray}
	[\omega^{2}\delta_{\mu\nu}
	+2i\omega~\epsilon_{\mu\nu\delta} \Omega_{\delta} 
	-C_{\mu\nu}({\bf k},\omega)]
	\xi_{{\bf k}\nu}(\omega)
	=0,
	\label{ev-equation}
\end{eqnarray}   
where, for the present non-retarded limit we define,
\begin{eqnarray}
%
	D_{\mu\nu}({\bf k})=\frac{1}{m_{\rm d}V_{dD}}
	\sum_{\bf q}q_{\mu}q_{\nu}\psi(q)[S(|{\bf k-q}|)-S(q)].
\end{eqnarray}   
Allowing us to write,
\begin{eqnarray}
	C_{\mu\nu}({\bf k},\omega)
	=\omega_{\rm pd}^{2}
	\left[\frac{k_{\mu}k_{\nu}}{k^{2}+\kappa_{b}^{2}}
	+\cal{D}_{\mu\nu}({\bf k},\omega)\right].
\end{eqnarray}   
where the dimensionless form of $D_{\mu\nu}$ is used, 
\begin{eqnarray}
	{\cal D}_{\mu\nu}({\bf k})&=&\sum_{\bf q}
	\frac{q_{\mu}q_{\nu}}{q^{2}+\kappa_{b}^{2}}
	[\tilde{S}(|{\bf k -q}|)-\tilde{S}(q)].
\end{eqnarray}   
In non-retarded limit, the only surviving transverse mode is via a transverse 
shear. This is indeed recovered from ${\cal D}_{\mu\nu}(\bf k)$ which contains 
both 
longitudinal and transverse shear. 
The longitudinal component of ${\cal D}_{\mu\nu}$ is 
recovered by operating ${\cal D}_{\mu\nu}({\bf k})$ over with longitudinal 
projection tensor ${\cal L}_{\mu\nu}({\bf k})={k_{\mu}k_{\nu}}{/k^{2}}$, 
\begin{eqnarray}
	{\cal D}_{L}({\bf k})&=&\sum_{\bf q}
	{\cal L}_{\mu\nu}({\bf k})
	\frac{q_{\mu}q_{\nu}}{q^{2}+\kappa_{b}^{2}}
	[\tilde{S}(|{\bf k -q}|)-\tilde{S}(q)],
\end{eqnarray}   
or,
\begin{eqnarray}
	{\cal D}_{L}({\bf k})&=&\sum_{\bf q}
	\frac{({\bf k\cdot q})^{2}}{k^{2}(q^{2}+\kappa_{b}^{2})}
	[\tilde{S}(|{\bf k -q}|)-\tilde{S}(q)].
\end{eqnarray}   
%
%
Similarly, the transverse component of ${\cal D}_{\mu\nu}$ is 
recovered by operating it over by the transverse
projection tensor 
$\frac{1}{2}{\cal T}_{\mu\nu}({\bf k})=\frac{1}{2}(\delta_{\mu\nu}-k_{\mu}k_{\nu}/k^{2})$,
%
\begin{eqnarray}
	\nonumber
	{\cal D}_{T}({\bf k})=\sum_{\bf q}
	\frac{1}{2}\{\delta_{\mu\nu}-{\cal L}_{\mu\nu}({\bf k})\}
	\frac{q_{\mu}q_{\nu}}{q^{2}+\kappa_{b}^{2}}\\
	\times [\tilde{S}(|{\bf k -q}|)-\tilde{S}(q)].
\end{eqnarray}   
Since, the contribution in transverse component comes only from non-diagonal 
terms this can be readily written as,
\begin{eqnarray}
	{\cal D}_{T}({\bf k})=\sum_{\bf q}
	-\frac{1}{2}{\cal L}_{\mu\nu}({\bf k})
	\frac{q_{\mu}q_{\nu}}{q^{2}+\kappa_{b}^{2}}
	[\tilde{S}(|{\bf k -q}|)-\tilde{S}(q)].
\end{eqnarray}   
In 3D conditions, the relation between ${\cal D}_{L}({\bf k})$ and 
${\cal D}_{T}({\bf k})$ becomes \cite{Golden},
\begin{eqnarray}
	{\cal D}_{T}({\bf k})&=&-\frac{1}{2}{\cal D}_{L}({\bf k}).
\end{eqnarray}   
We now find the dispersion relation by allowing the determinant to vanish
for nontrivial solutions \cite{golden1993dielectric}, 
\begin{eqnarray}
	\det[\omega^{2}\delta_{\mu\nu}
	+2i\omega~\epsilon_{\mu\nu\delta} \Omega_{\delta} 
	-C_{\mu\nu}({\bf k},\omega)]
	=0,
	\label{ev-equation}
\end{eqnarray}   
which, up on substituting the elements of matrix $C_{\mu\nu}$ and 
considering ${\bf \Omega}=\Omega\hat{z}$, becomes,
%
%
\begin{eqnarray}
%
%
%
%
	\nonumber
	\left[\omega^{2}-
	\left\{\omega_{0}^{2}({\bf k})+D_{L}({\bf k})\right\}\right]
	\left[\omega^{2}-
	D_{T}({\bf k})\right]\\
	-\omega^{2}(2\Omega)^{2}=0,
	\label{solution-QE}
\end{eqnarray}   
where, $\omega_{0}^{2}({\bf k})=\omega_{\rm pd}^{2}{k^{2}}{/(k^{2}+\kappa_{b}^{2})}$,
such that solutions of Eq.~(\ref{solution-QE}) are,
\begin{eqnarray}
%
	\nonumber
	\omega^{2}({\bf k})=\frac{1}{2}[
	\{\omega_{0}^{2}({\bf k})+D_{L}({\bf k})+D_{T}({\bf k})\}+(2\Omega)^{2}]\\
	\nonumber
	\pm \frac{1}{2}\left(
	[\{\omega_{0}^{2}({\bf k})+D_{L}({\bf k})+D_{T}({\bf k})]
	+(2\Omega)^{2}\}^{2}\right.\\
	\left.-4D_{T}({\bf k})
	[\omega_{0}^{2}({\bf k})+D_{L}({\bf k})]\right)^{1/2},
%
%
	\label{general-dispersion}
\end{eqnarray}   
yielding, in the case $\Omega\rightarrow 0$, the well known frequencies 
for longitudinal and transverse modes, 
\begin{eqnarray}
	\omega_{L}({\bf k})
	&=&[\omega_{0}^{2}({\bf k})+D_{L}({\bf k})]^{1/2}
	\label{longitudinal}
	\\
	\omega_{T}({\bf k})&=&[D_{T}({\bf k})]^{1/2}.
%
	\label{transverse}
\end{eqnarray}   
It is now possible to examine the general QLCA dispersion 
(\ref{general-dispersion}) which accommodates all the essential effects
for this setup, namely, the strong coupling ($D_{\mu\nu}\neq 0$),
the Yukawa potential and the dust rotation.
The strong coupling effects can be included in their simplest form by adopting 
the long wavelength limit $\Omega a/c\ll ka \ll 1$, or the ``excluded volume 
consideration'' \citep{khrapak2016long} where $D_{L,T}$ take a relatively approximate but simpler 
analytic form, given by,
\begin{eqnarray}
	D_{L,T}({\bf k})=\omega_{\rm pd}^{2}\int_{0}^{\infty}dr
	\frac{e^{-\kappa r}}{r}
	[g(r)-1]{\cal K}_{L,T}(kr,\kappa r), 
	\label{D-long-wave-length}
\end{eqnarray}   
where,
\begin{eqnarray}
	\nonumber
	{\cal K}_{L}(kr,\kappa r)=
	-2\left[1+\kappa r+\frac{1}{3}(\kappa r)^{2}\right]\\
	\nonumber
	\times\left[
	\frac{\sin{kr}}{kr}
	+3\frac{\cos{kr}}{(kr)^2}
	-3\frac{\sin{kr}}{(kr)^3}
	\right]\\
	+\frac{1}{3}(\kappa r)^2
	\left(
	\frac{\sin{kr}}{kr}-1
	\right),
	\label{def-KL}
\end{eqnarray}   
and
\begin{eqnarray}
	\nonumber
	{\cal K}_{T}(kr,\kappa r)=
	\left[1+\kappa r+\frac{1}{3}(\kappa r)^{2}\right]\\
	\nonumber
	\times\left[
	\frac{\sin{kr}}{kr}
	+3\frac{\cos{kr}}{(kr)^2}
	-3\frac{\sin{kr}}{(kr)^3}
	\right]\\
	-\frac{1}{3}(\kappa r)^2
	\left(
	\frac{\sin{kr}}{kr}-1
	\right).
	\label{def-KT}
\end{eqnarray}   
The function $g(r)$ in this limit is approximated to a step-like function, 
having uniform strength 1
in the region beyond a minimum radius $R$ and remaining zero within 
the region $r<R$, allowing longitudinal and transverse mode frequencies to be 
recovered, respectively, as, 
%
%
\begin{eqnarray}
 \nonumber
	\omega_{L}^{2}= (2\Omega)^{2}+\omega_{\rm pd}^{2}e^{-k R}\left[
	(1+k R)\left(\frac{1}{3}-\frac{2\cos{kR}}{k^2R^2}
	\right.\right.\\
	\left.\left.
	+\frac{2\sin{kR}}{k^3R^3}\right)
	-\frac{\kappa^{2}}{\kappa^{2}+k^{2}}
	\left(\cos{kR}+\frac{\kappa}{k}\sin{kR}\right)  \nonumber
	\right.\\
	\left.
+(1+k R)\left(\frac{1}{3}-\frac{\cos{kR}}{(kR)^2}+\frac{\sin{kR}}{(kR)^3}\right) \nonumber
\right.\\
\left.
\times \frac{(2\Omega)^{2}}{(2\Omega)^{2} + \omega^{2}_{0}} \right],
\end{eqnarray} 
and
%
%
\begin{eqnarray}
	\omega_{T}^{2}=\frac{\omega^2_{0} \omega_{\rm pd}^{2}e^{-k R}
	(1+k R)}{(2\Omega)^{2} + \omega^2_{0}}\left[\frac{1}{3}-\frac{\cos{kR}}{(kR)^2}
	+\frac{\sin{kR}}{(kR)^3}\right].
\end{eqnarray}

\begin{figure}[hbt]
 \centering
 \includegraphics[width=90mm]{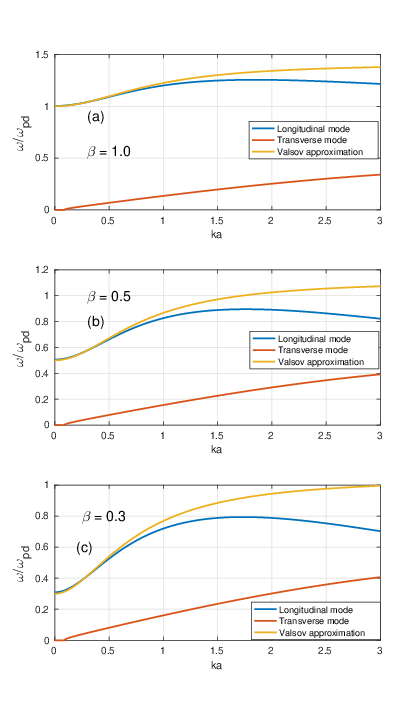}
	\caption{Longitudinal and transverse dispersion relations $\omega_{L}(k)$ and $\omega_{T}(k)$ plotted in (a), (b) and (c) for varying strength of rotation $\beta$=1, 0.5 and 0.3, respectively.}
\label{dispersion-LT}
\end{figure}

These dispersion relations are plotted in Fig.~\ref{dispersion-LT}, along with 
the RPA (or Vlasov) approximated version of $\omega_{L}(k)$. 
For computing $\omega_{L,T}(k)$ plotted in Fig.~\ref{dispersion-LT}, we have 
used $\kappa=1$, $\Gamma=180$, and $R=1+\kappa/10$ approximately, 
following \citep{khrapak2016long,khrapak2017practical}, while using $\beta$ 
as a variable having values 0.3 - 1.0 in order to account for the additional 
effect of rotation present in our expressions.
One of the most notable features of $\omega_{L,T}(k)$ is the reducing difference 
between the saturated values of RPA approximated version of $\omega_{L}(k)$ 
and its strongly coupled version, with increasing $\Omega$. The latter 
essentially oscillates about a saturation value as generally prescribed by 
the QLCA formulation.
This similarity of the solutions at stronger rotation is discussed further in
the context of experimental results which are compared to RPA limit dispersion 
in Sec.~\ref{solutions}B and show good agreement.
As another notable feature, an onset exists in the transverse mode frequency 
with respect to small $k$ values where the first non-zero frequency corresponds 
to a minimum value of $k$.
This onset was recovered and discussed in larger detail in the recent 
important studies of QLCA dispersion in Yukawa systems \citep{khrapak2019onset}. 
%
%
Before characterizing the solutions, we briefly show that the RPA limit is 
indeed analogous to the fluid limit by obtaining the same results 
using a multi-component fluid model, or in the 
weakly coupled limit, $D_{L,T}\rightarrow 0$, of the QLCA formulation.
The applied multi-component fluid model is used also for producing numerical 
solutions, in order to generate independent validation of the RPA results 
in Sec.~\ref{solutions}B.
\subsection{The corresponding multi-component fluid model}
We now use the set of 2D fluid equations, in a rotating non-inertial 
frame of reference, in order to model the weakly coupled rotating dusty 
plasma where the Coriolis force takes over the role of the Lorentz force. 
Here, the dust fluid equilibrium is assumed to be a rigid body like rotation
of the dusty plasma with an angular frequency ${\bf \Omega}$ pointing along 
$\hat{z}$. The components of momentum 
equation for rotating isothermal dusty fluid in the x-y plane then have 
the following form,     
 \begin{eqnarray}
	 \frac{\partial{u}_{{\rm d}x} }{\partial t} + ({\bf u}_{{\rm d}}
	 \cdot\nabla) {u}_{{\rm d}x} 
	 = \frac{q_{\rm d}}{m_{\rm d}}{E}_{x} 
	 + 2{u}_{dy}{\Omega},\\ \label{eq:xmomentum}
	 \frac{\partial{u}_{{\rm d}y} }{\partial t} + ({\bf u}_{{\rm d}}
	 \cdot\nabla) {u}_{{\rm d}y} 
	 = \frac{q_{\rm d}}{m_{\rm d}}{E}_{y} 
	 - 2{u}_{{\rm d}x}{\Omega}, \label{eq:ymomentum}
\end{eqnarray}
where $u_{dx}$ and $u_{dy}$ are the dust velocity along x and y-direction, 
respectively, $q_{\rm d} = -Z_{\rm d}e$ is the dust charge and the 
electric field 
${\bf E} = -\nabla\phi$.
The centrifugal force is once again assumed to be balanced by the force 
due to the confining potential with the angular frequency ${\bf \Omega}$ 
of the rotation radially uniform.
The continuity equation for the dust is, 
 \begin{equation}
  \frac{\partial n_{\rm d}}{\partial t} 
	 + \nabla\cdot n_{d0} \textbf{u}_{\rm d} = 0.
  \label{eq:continuity}
 \end{equation} 
For the low frequency wave phenomena of the dust medium, the 
background ions and electrons densities are well model by the Boltzmann 
relation,
 \begin{eqnarray}
	 n_{\rm e} = n_{{\rm e}0} \exp\left({\frac{e \phi}{k_{B}T_{\rm e}}}\right), 
	 \\\label{eq:e-boltzmann}
	 n_{\rm i} = n_{{\rm i}0} \exp\left({\frac{-e \phi}{k_{B}T_{\rm i}}}\right).
\label{eq:i-boltzmann}
\end{eqnarray}
The set of equations is completed by the Poisson equation, 
 \begin{equation}
  \frac{\partial^2\phi}{\partial x^2} = -4\pi e\left[n_{\rm i}-n_{\rm e}-Z_{\rm d}n_{\rm d}\right],
\label{eq:poisson_equation}
\end{equation}
with the following equilibrium charge neutrality condition in this three 
component plasma, 
\begin{equation}
	n_{{\rm i}0}e =-n_{{\rm e}0}e-Z_{\rm d}en_{{\rm d}0}.
	\label{eq:charge_neutrality}
\end{equation}
%
For a small amplitude, plane-wave-like, pure electrostatic perturbation, 
such that ${\bf k \| u}$, combining the linear form of 
Eq.~(\ref{eq:xmomentum})-(\ref{eq:charge_neutrality}) readily yields 
the same dispersion relation as 
Eq.~(\ref{longitudinal}), 
\begin{equation}
\omega^2 = \left(2\Omega\right)^2 
	+ \left[\frac{\omega^2_{\rm pd}}{1
	+{(k \lambda_{\rm i})^{-2}} 
	+{(k \lambda_{\rm e})^{-2}}}
	\right],
	\label{eq:Dispersion_relation}
\end{equation} 
although limited to response of dust to only the longitudinal (compressible) 
component of the initial perturbation because of the choice ${\bf k \| u}$. 
Note that an independently propagating pure transverse perturbation is only 
realizable in $\Omega\rightarrow 0$ limit and is ruled out in the presence 
of rotation as the transverse and longitudinal velocity components are 
coupled by the rotation.
Recovery of dispersion 
(\ref{eq:Dispersion_relation}) shows that the RPA limit of the QLCA 
formulation suitably represents the fluid limit of the dust response, 
also for the rotating dusty plasma system, considered in the above general 
treatment.
%
\section{Rotating dusty plasma dispersion in RPA limit and its
numerical validation \label{solutions}}
In order to obtain the wave-like solutions of rotating dusty fluid 
perturbations, governed by the linear form of 
Eq.~(\ref{eq:xmomentum})-(\ref{eq:charge_neutrality}), 
we have used the pseudo-spectral technique for two dimensional solutions in 
(x-y) plane, perpendicular to the direction of the angular frequency 
${\bf \Omega}$ in pure Cartesian geometry. The solutions thus represent 
excitations obtained in a frame co-rotating with the dusty plasma, hence 
can be suitably compared with the dispersion relations 
(\ref{longitudinal})
and (\ref{eq:Dispersion_relation}) derived in the 
same non-inertial frame of reference.
%
\subsection{Numerical scheme and normalization}
In our numerical procedure the spatial and temporal discretisation are done
using $k$ and $t$ variables, respectively, and
satisfy the Courant-Friedrichs-Lewy (CFL) condition. 
The Predictor-corrector method has been 
used for time-stepping. For all the results presented in the following 
numerical analysis we choose a 2-dimensional grid of the size 
of $N_{x}\times N_{y}$ 
= 128$\times$128. The spatial resolution is determined by the 
combination of grid size and the system length 
while the typical time-step value chosen is $\delta$t = $10^{-3}$ 
$\omega_{\rm pd}^{-1}$. Amplitudes of the perturbation in velocities 
components and density are all chosen to be the equal with values 
$\delta u_{\rm d}/\lambda_{\rm D}\omega_{\rm pd}$ = 
$\delta n_{\rm d}/n_{{\rm d}0}$ = 0.001.

The normalizations chosen for the numerical computation of the 
solutions, as well as for the solutions presented in this and following 
sections, are as follows.
%
Relevant to most dusty plasma system, the time and length values are 
normalized to the inverse dust acoustic frequency $\omega_{\rm pd}^{-1}$ and 
mean dust particle separation $a$,
respectively,
As a result, the rotation frequency and velocity have the normalizations
$\omega_{\rm pd}$ and 
$a\omega_{\rm pd}$, 
respectively.
We additionally define a parameter $\beta=2\Omega/\omega_{\rm pd}$ denoting 
the strength of the rotation, or alternatively, the strength of the 
Coriolis force acting on the dust in the frame co-rotating with the dust.
The another most important parameter remains the screening parameter, or the 
ratio of the mean dust particle separation to the effective Debye length,
$\kappa=a/\lambda_{\rm D}$. Since $\kappa_{b}=1/\lambda_{\rm D}$,
we get, $\kappa=\kappa_{b}a$.

\subsection{Collective dust mode dispersion in the rotating frame}
Wave-like solutions are obtained form the numerical evolution of 
the dust parameter perturbations that follows the 
equations~(\ref{eq:xmomentum})-(\ref{eq:charge_neutrality}), 
with periodic boundary condition which are implemented along both 
$x$ and $y$-directions.
The numerical dispersion relations are constructed 
by simulating the above evolution for a range of wave vector values.
Detailed parametric characterization of the obtained dispersion is mainly 
done by variation in the values of two key parameters, $\beta$ and $\kappa$. 
The parameter $\beta$ represents the strength of Coriolis force and 
$\kappa$ is screening parameter, representing the characteristic length 
of the background plasma generated screening scaled to the mean 
dust particle separation $a$. 
\begin{figure}[hbt]
 \centering
 \includegraphics[width=90mm]{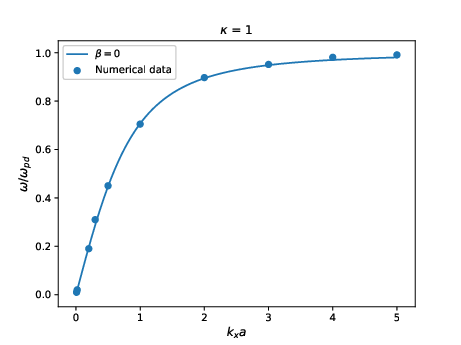}
 \caption{ Analytical (line) and numerical (points)  longitudinal  dispersion relations  plotted with screening parameter $\kappa$ = 1 and without rotation.}  
\label{fig1}
\end{figure}
The present analysis is done using one-dimensional propagation of the wave, 
by choosing 
$\textbf{k}$ = $k\hat{x}$ such that $\textbf{E}$ = $E_{x}\hat{x}$ and 
the only the longitudinal perturbations are recovered. 
%
In order to analyze solution regime relevant to existing experiments 
incorporating dust rotation \cite{hartmann2013}, for the present 
computations we have used 
$\Omega$ $\approx$ 23.3 to 25 rad/s and 
$\omega_{\rm pd}$ $\approx$ 66 to 106 rad/s, 
which translates in the range of parameter $\beta$ $\approx$ 0.4 to 0.8.
The rest of the parameters are also chosen to have their experimental value.
Accordingly, $m_{\rm d}$ = 6.64$\times$ $10^{-14}$ kg, 
$q_{\rm d}$ = 6300$e$. Similarly the value of parameter $\kappa$ is also used
as in the experiment for the corresponding cases. The computations 
are done for the three cases, obtaining the dispersion relation as plotted 
in figures presented further below in this section.

As the first, reference case, Fig.~$\ref{fig1}$ shows the 
longitudinal wave dispersion relation in the absence of rotation 
($\Omega$ = 0). This represents the dispersion relation for a regular 
dust acoustic wave in the laboratory frame with frequency saturating 
into the value $\omega=\omega_{\rm pd}$ for large $k$ values, or 
in the limit 
$ka\gg 1$.
The dots in the Fig.~\ref{fig1} represent the simulated 
value, whereas the solid line is the analytic dispersion relation. 
\begin{figure}[hbt]
 \centering
 \includegraphics[width=90mm]{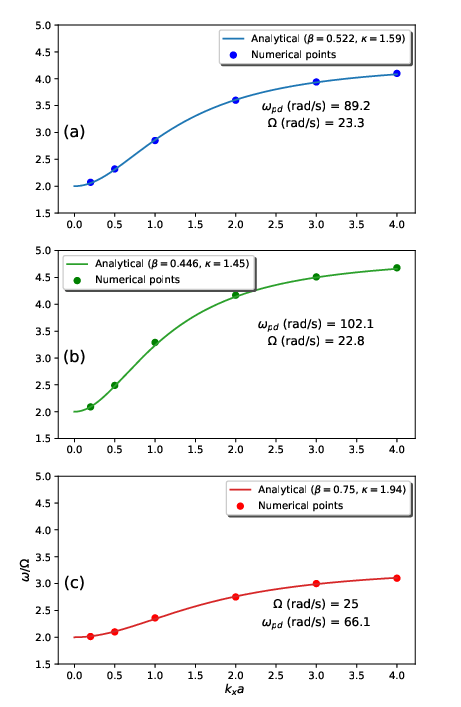}
 \caption{ Analytical (lines) and numerical (points)  longitudinal dispersion relations  plotted with (a) $\Omega$ = 23.3 rad/s, $\omega_{pd}$ = 89.2 rad/s and $\kappa$ = 1.54, (b) $\Omega$ = 22.3 rad/s, $\omega_{pd}$ = 102.1 rad/s and $\kappa$ = 1.45, and (c) $\Omega$ = 25 rad/s, $\omega_{pd}$ = 66.3 rad/s and $\kappa$ = 1.94.} 
\label{fig2}
\end{figure}
Results plotted in Fig.~$\ref{fig2}$ show the longitudinal wave dispersion 
relation for a finite values of $\Omega$ and therefore they indeed correspond 
to transformed data from the laboratory measurements which are obtained in a 
rotating frame. 
The sets of parameters for producing the cases in Fig.~\ref{fig2} are
derived from the experimental conditions of Ref.~\citep{hartmann2013}, 
where dispersion data was measured over a range of $\beta$ and $\kappa$ 
values in the frame of reference co rotating with the dust cloud.
Similar to Fig.~\ref{fig1}, the dots in the Fig.~\ref{fig2}(a)-\ref{fig2}(c) 
represent the simulated value, whereas the solid lines are the analytic 
dispersion relations, plotted here using the same set of parameters as 
used in the corresponding experimental cases.
The frequency values on the entire $ka$ range for all the three cases 
presented in Fig.~\ref{fig2}(a)-\ref{fig2}(c) closely agree with the 
experimental measurements. 

In more specific terms, the dispersion relation plot presented in 
Fig.~\ref{fig2}(a) corresponds to the set of parameters, $\Omega$ = 23.3 
rad/s, $\omega_{pd}$ = 89.2 rad/s (i.e.$\beta=0.522$) and $\kappa$ = 1.54.  
The dispersion at value $k\rightarrow 0$ is found to start duly from frequency
2$\Omega=46.6$ rad/s. At low wave numbers, the excitation frequency increases 
with the wave number which agree with the experimental results 
\citep{hartmann2013} and with increasing $k$-value, depending upon $\beta$, 
wave dispersion relation attains a saturated value which also confirms with 
what measured in the experiment. 
Similarly, the dispersion relation plotted in Fig.~\ref{fig2}(b) 
is also simulated using the set of parameters, $\Omega$ = 22.3 rad/s, 
$\omega_{pd}$ = 102.1 rad/s (i.e., $\beta=0.466$) and $\kappa$ = 1.45, 
and shows the same trend when compared to experimental data. 
Note, however, that although all dispersion curves start with frequency
which is two times the rotation frequency at ${k\rightarrow 0}$, their 
height, or the saturated value at ${k\gg 1}$ are different in each case, 
and increase in this saturated value is also in close agreement with the 
experimental observation for the corresponding. 
The last set plotted for comparison with experiments in Fig~\ref{fig2}(c) 
corresponds to the set of parameters using stronger rotation, 
$\Omega$ = 25 rad/s, $\omega_{pd}$ = 66.3 rad/s (i.e., $\beta = 0.75$) 
and $\kappa$ = 1.94. 
The saturation value of the frequency for this set is found to be minimum 
among the three cases presented. The dispersion curves therefore show a 
tendency to flatten and approach a minimum value as the value of parameter 
$\beta$ is increased.
This limit is discussed further in the additional cases presented for a 
detailed analysis, independent of experiments, further below.

The overall agreement with the experimental observations 
is found better for the cases with higher 
$\beta=2\Omega/\omega_{\rm pd}$. For example, the saturation value of the
present RPA results plotted for the third set of parameters in 
Fig.~\ref{fig2}(c), as well as that in the corresponding experimental case, 
is $\omega/\Omega\sim 3$, which is in very close agreement with each other. 
This difference between the two however grows for smaller
$\Omega$, i.e., in the cases plotted in Fig.~\ref{fig2}(a) and \ref{fig2}(b) 
where the RPA results saturate at slightly higher values than the 
corresponding experimental levels. As an additional aspect, the saturation
in the $\omega({\bf k})$ is attained when full 3-dimensional form
of the potential, $\psi_{dD}\equiv \phi_{3D}$ is accounted for. While 
using the exact 2-dimensional form for $\psi_{dD}$ does not produce 
saturation in $\omega({\bf k})$ \cite{jiang2007theoretical}, 
the 3-dimensional form applies when the axial (along $\hat{z}$) variation in 
the dust cloud is rather week because of boundaries in $\hat{z}$ direction 
are sufficiently distant creating a nearly 3-dimensional dust cloud. 
The above agreement therefore additionally highlights enhanced 3-dimensional 
attributes of the dust dynamics in a rotating dust experiment, and, in turn, a
subdued impact of the sedimentation usually produced by gravity in typical
laboratory dusty plasmas that are levitated by an upward directed sheath 
potential of a horizontally placed electrode.
\subsection{Longitudinal-Transverse coupling by dust rotation}
The two dimensional numerical simulations on an extended 2-D grid allows us to
illustrate the inevitable coupling between the longitudinal and transverse 
excitations and emergence of compression in an initially pure transverse 
shear like (non-compressional) perturbation of the 2-D velocity field. 
This effect readily follows from the solution (\ref{ev-equation}) which 
is reducible in two independent longitudinal and transverse dispersion 
relations, (\ref{longitudinal}) and (\ref{transverse}), only in the limit 
$\Omega\rightarrow 0$.
\begin{figure}[hbt]
 \centering
 \includegraphics[width=95mm]{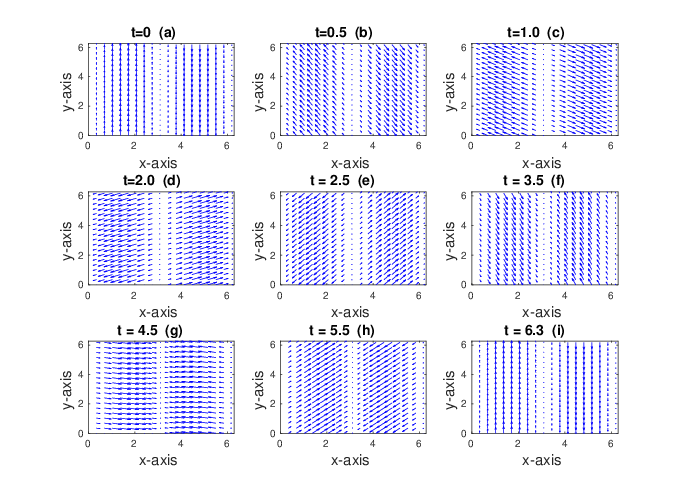}
	\caption{ 2D velocity vector field profiles at different time for $\kappa\rightarrow \infty$ and $\beta=1$ ($\Omega=\omega_{\rm pd}/2)$, i.e., the velocity vector complets one rotation in half the system rotation.}
 \label{velocity-field}
\end{figure}
Illustrating this coupling between the two modes in a rotating dust setup, 
the total velocity vector field is plotted in Fig.~\ref{velocity-field} at 
different phases (subperiodic time values) of an initially divergence free 
velocity perturbation as visible in Fig.~\ref{velocity-field}(a). 
The evolution is presented for $\kappa\rightarrow \infty$ and $\Omega>0$ case when
no electric field is expected because of perfectly shielded dust.
The periodic evolution of the velocity field through 
Fig.~\ref{velocity-field}(a)-Fig.~\ref{velocity-field}(i) displayes 
emergence of divergent fields at the intermediate phases (e.g., t=2.0, 4.5 $\omega_{\rm pd}^{-1}$). 
Consequently, in the general case of an imperfectly shielded dust, finite
electrostatic field would still develop, even if the initial perturbations 
are chosen to be purely divergence-free, ruling out a pure shear wave.
\subsection{General characterization with respect to $\beta$ and $\kappa$}
\begin{figure}[hbt]
 \centering
 \includegraphics[width=90mm]{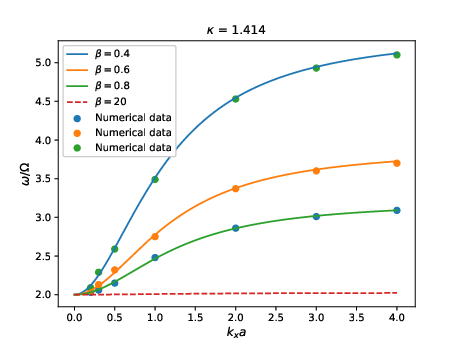}
  \caption{ Analytical (lines) and numerical (points)  longitudinal wave dispersion relations plotted with different value of $\beta$ and constant $\kappa$ = 1.41.}
\label{fig3}
\end{figure}

\begin{figure}[hbt]
 \centering
 \includegraphics[width=90mm]{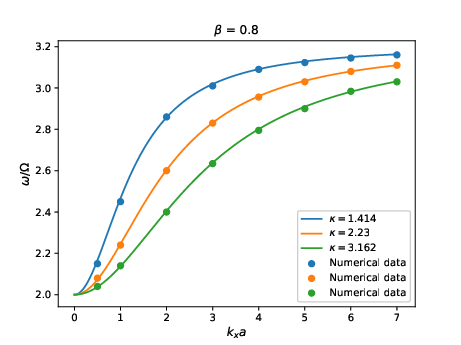}
  \caption{ Analytical (lines) and numerical (points) longitudinal wave dispersion relations plotted with different value of $\kappa$ and constant $\beta$ = 0.8 .}
 \label{fig4}
\end{figure}

Independent of their comparison with experimental data sets,
a more systematic additional analysis is presented in Fig.~\ref{fig3} and 
Fig.~\ref{fig4} which are obtained by simulating the cases with exclusive
variation in the parameters $\beta$ and $\kappa$, respectively.

The value of $\beta=0.4$, 0.6 and 0.8 are used for the dispersion relations
presented in Fig.~\ref{fig4}(a) with a fixed value of $\kappa=1.41$.
The increase in strength of rotation or $\beta$ reduces the rise of wave 
frequency value from the rotational frequency $\Omega$ and also the 
saturated value of the wave frequency. This tendency indicates
the dominance of the Coriolis force effect over the electrostatic force 
effects, associated with dust acoustic mode, in the original equation of 
motion (\ref{particle-eq}) written in the rotating frame.
In order to highlight the behavior of the dispersion curve in the limiting 
case of very high rotation, dispersion curve is presented for a very high 
$\beta$ value (dashed line) showing that the wave frequency can be almost 
independent of $k$ and nearly equal to $2\Omega$ for very large $\beta$.

The impact of screening parameter $\kappa=a/\lambda_{\rm D}$ on the 
mode frequency is similarly analyzed in Fig.~\ref{fig4}(b) where 
dispersion curve with values of $\kappa=1.41$, 2.23 and 3.16 and a fixed value 
of $\beta=0.8$, are plotted, respectively. Increase in $\kappa$ value, 
or high dust density, causes an early saturation of the dispersion curve 
with respect to $ka$. While the saturated value of the frequency remains
independent of $\kappa$, at very small screening parameter the approach to 
the saturated value become very slow with increasing $ka$.

\subsection{Approach to weak coupling like effects by rotation}
While the QLCA expressions for frequencies apply to general cases, 
including the effects of strong coupling, the numerical solutions presented
and compared with the experimental results are from the fluid model corresponding to the random phase approximation (RPA) limit of the QLCA formulation. 
The observed close correspondence between this the weakly coupled regime of 
the model to somewhat contrasting considerations of strongly coupled limit  
of the dust particles as understood in the rotating plasma experiment 
\citep{hartmann2013} motivates us to explore the possible explanations.  
Especially as the correspondence is seen to improve between the two for
higher rotation frequency, as discussed further below.
Two aspects can be discussed in this respect
reconciling the agreement of the experimental data with the weak coupling 
(RPA) results rather than with the branches corresponding to the strong 
coupling for which $D_{\mu\nu}$ remains finite. The first argument uses the 
non-equilibrium thermodynamic state of dust medium which is driven 
externally by an applied rotation. 
Considering that the dust medium is not in its thermal equilibrium, perhaps a 
modification in its coupling strength description can be explored in order 
to accommodate the parameter other than temperature in quantifying its 
effective coupling. 
A nonequilibrium state is no longer described by a single state parameter. 
Since the equilibrium ensembles yield the dust temperature as the unique 
parameter describing its thermodynamic state, this remains the only variable 
determining the degree of coupling in the dust for non rotating conditions, 
or even a magnetized dusty plasma. Rotating dust, on the other hand, being an 
open system and therefore operating away from the thermodynamic equilibrium 
presents a case, distinct from both non-rotating as well as magnetized dusty 
plasma that are essentially in a thermodynamical equilibrium.

An alternate argument relates to admitting only a small variation in its 
thermodynamical state from the equilibrium state, but attributing it to a 
possible rise, introduced by the external rotation, in the internal energy 
of the dust medium that results in its effective transition to a well defined 
weakly coupled, or fluid-like, regime under the standard, equilibrium 
formulation.
This argument is also supported by the clearly measured changes in the 
characteristic parameters of the dust when rotation is switched on. 
For example, shown using a four particle dust cluster by K\"{a}hlert et al.\citep{kahlert2012} 
that as compared to the non-rotating case, the system is closer to the liquid 
state and has a lower Coulomb coupling.
In a harmonic potential with confinement frequency $\omega_{\perp}$, the
coupling parameter in rotating and non-rotating configurations, scale as
$\Gamma(\Omega)/\Gamma(0)=(1-\Omega^{2}/\omega_{\perp}^{2})^{1/3}$,
while the screening parameter scales as,
$\kappa(\Omega)/\kappa(0)=(1-\Omega^{2}/\omega_{\perp}^{2})^{-1/3}$.
We indeed note that the agreement with the experimental observations 
is improved significantly for the cases with higher 
$\beta=2\Omega/\omega_{\rm pd}$. For example, the saturation value of the
present RPA results plotted for the third set of parameters in 
Fig.~\ref{fig2}(c), as well as that in the corresponding experimental case, 
is $\omega/\Omega\sim 3$, which are therefore in very close agreement with 
each other. 
This agreement however becomes slightly poorer at 
weaker rotation frequency $\Omega$, or for small $\beta$ as in the cases 
plotted in Fig.~\ref{fig2}(a) and \ref{fig2}(b) where the RPA results
predict saturation at slightly higher values than the experiment. 
It should be further noted that in the present formulation treating
a stably rotating dust fluid,
the confining 
frequency $\omega_{\perp}$ is considered to be suitably high such that 
rotations with large enough $\Omega$ are admissible and $\Gamma$ and 
$\kappa$ are nearly independent of $\Omega/\omega_{\perp}$ \cite{kahlert2012}. 
In contrast, the change in experimental dust parameters with respect to 
$\Omega$ is much stronger for the weak coupling effect to be more 
clearly visible at larger $\Omega$. 

Both these scenarios however highlight a distinction between the underlying
thermodynamical characters of a dusty plasma which is subjected to a magnetic 
field and the one which is subjected to an externally enforced rotation.  
The latter transforming its thermodynamical state to an open 
system, therefore no longer being governed by an equilibrium formulation. 
\section{Summary and conclusion}
In the treatment presented we worked out a QCLA formulation which 
explicitly accounts for dust rotation in a Yukawa system. Its 
limiting case is derived corresponding to the weakly coupled (or RPA) limit 
where it yields equivalence to the regular explicitly rotating ``dust fluid'' 
formulation. Finite rotation version of both strong and weak coupling case 
dispersion relation is derived and analyzed, showing correspondence between
`the faster rotating but weakly coupled' and the strongly coupled 
dust responses.
By presenting an equivalent multi-fluid rotating dusty plasma 
model and by means of its computational solutions for the parameters as 
in recent rotating dusty plasma experiments, independent agreement with 
analytical RPA dispersion relations is produced. A systematic characterization 
of the solutions with respect to variation in rotation frequency $\omega$ and 
screening parameter $\kappa$ is done separately.
The analytical and numerical analysis of the rotating plasma dispersion 
showed that if the weakly coupled limit is considered 
with distant axial boundaries (3D effects), the frequency saturation not 
only becomes achievable in the rotating dusty plasma but the saturated 
frequency values also show quantitative agreement with those recovered 
in the recent experiments of the rotating dusty plasmas. 
This behavior indicates that a dust medium with large enough coupling 
parameter $\Gamma$, when 
driven to a nonequilibrium state by an external rotation, not only displays 
weak coupling like effects but also a relaxation from the boundary effects 
such that the plasma bulk strongly limited by axial boundaries still displays 
nearly 3D characteristics. The rotation of the dust thus additionally 
appears to relax the 2D effects (2D sedimentation, usually generated by 
the gravity) on the dust cloud. 
It is therefore concluded that this nonequilibrium rotating system not only 
shows isomorphism with magnetized dust regime but perhaps also with its
microgravity regime, rendering the latter realizable also in the regular 
ground based laboratory experiments, if dust is subjected to a 
rotation. 
The analysis therefore motivates studies both in magnetized as well as in 
microgravity conditions in order to examine their similarity with 
ground based rotating dusty plasma experiments. The analysis may thus 
serve as a link 
between ground based magnetized dusty plasma experiments, like MDPX, and 
the International Space Station (ISS) based experiments in microgravity 
conditions. 
 \section{AIP PUBLISHING DATA SHARING POLICY}\label{SC}
 The data that support the findings of this study are available from the corresponding author upon reasonable request.
 

 \bibliographystyle{apsrev4-1}
 \bibliography{paper_2}

\begin{thebibliography}{30}%
\makeatletter
\providecommand \@ifxundefined [1]{%
 \@ifx{#1\undefined}
}%
\providecommand \@ifnum [1]{%
 \ifnum #1\expandafter \@firstoftwo
 \else \expandafter \@secondoftwo
 \fi
}%
\providecommand \@ifx [1]{%
 \ifx #1\expandafter \@firstoftwo
 \else \expandafter \@secondoftwo
 \fi
}%
\providecommand \natexlab [1]{#1}%
\providecommand \enquote  [1]{``#1''}%
\providecommand \bibnamefont  [1]{#1}%
\providecommand \bibfnamefont [1]{#1}%
\providecommand \citenamefont [1]{#1}%
\providecommand \href@noop [0]{\@secondoftwo}%
\providecommand \href [0]{\begingroup \@sanitize@url \@href}%
\providecommand \@href[1]{\@@startlink{#1}\@@href}%
\providecommand \@@href[1]{\endgroup#1\@@endlink}%
\providecommand \@sanitize@url [0]{\catcode `\\12\catcode `\$12\catcode
  `\&12\catcode `\#12\catcode `\^12\catcode `\_12\catcode `\%12\relax}%
\providecommand \@@startlink[1]{}%
\providecommand \@@endlink[0]{}%
\providecommand \url  [0]{\begingroup\@sanitize@url \@url }%
\providecommand \@url [1]{\endgroup\@href {#1}{\urlprefix }}%
\providecommand \urlprefix  [0]{URL }%
\providecommand \Eprint [0]{\href }%
\providecommand \doibase [0]{http://dx.doi.org/}%
\providecommand \selectlanguage [0]{\@gobble}%
\providecommand \bibinfo  [0]{\@secondoftwo}%
\providecommand \bibfield  [0]{\@secondoftwo}%
\providecommand \translation [1]{[#1]}%
\providecommand \BibitemOpen [0]{}%
\providecommand \bibitemStop [0]{}%
\providecommand \bibitemNoStop [0]{.\EOS\space}%
\providecommand \EOS [0]{\spacefactor3000\relax}%
\providecommand \BibitemShut  [1]{\csname bibitem#1\endcsname}%
\let\auto@bib@innerbib\@empty
\bibitem [{\citenamefont {Yaroshenko}\ \emph {et~al.}(2007)\citenamefont
  {Yaroshenko}, \citenamefont {Verheest},\ and\ \citenamefont
  {Morfill}}]{yaroshenko2007}%
  \BibitemOpen
  \bibfield  {author} {\bibinfo {author} {\bibfnamefont {V.}~\bibnamefont
  {Yaroshenko}}, \bibinfo {author} {\bibfnamefont {F.}~\bibnamefont
  {Verheest}}, \ and\ \bibinfo {author} {\bibfnamefont {G.}~\bibnamefont
  {Morfill}},\ }\href@noop {} {\bibfield  {journal} {\bibinfo  {journal}
  {Astronomy \& Astrophysics}\ }\textbf {\bibinfo {volume} {461}},\ \bibinfo
  {pages} {385} (\bibinfo {year} {2007})}\BibitemShut {NoStop}%
\bibitem [{\citenamefont {Yaroshenko}\ \emph {et~al.}(2009)\citenamefont
  {Yaroshenko}, \citenamefont {Ratynskaia}, \citenamefont {Olson},
  \citenamefont {Brenning}, \citenamefont {Wahlund}, \citenamefont {Morooka},
  \citenamefont {Kurth}, \citenamefont {Gurnett},\ and\ \citenamefont
  {Morfill}}]{Yaroshenko}%
  \BibitemOpen
  \bibfield  {author} {\bibinfo {author} {\bibfnamefont {V.}~\bibnamefont
  {Yaroshenko}}, \bibinfo {author} {\bibfnamefont {S.}~\bibnamefont
  {Ratynskaia}}, \bibinfo {author} {\bibfnamefont {J.}~\bibnamefont {Olson}},
  \bibinfo {author} {\bibfnamefont {N.}~\bibnamefont {Brenning}}, \bibinfo
  {author} {\bibfnamefont {J.-E.}\ \bibnamefont {Wahlund}}, \bibinfo {author}
  {\bibfnamefont {M.}~\bibnamefont {Morooka}}, \bibinfo {author} {\bibfnamefont
  {W.}~\bibnamefont {Kurth}}, \bibinfo {author} {\bibfnamefont
  {D.}~\bibnamefont {Gurnett}}, \ and\ \bibinfo {author} {\bibfnamefont
  {G.}~\bibnamefont {Morfill}},\ }\href@noop {} {\bibfield  {journal} {\bibinfo
   {journal} {Planetary and Space Science}\ }\textbf {\bibinfo {volume} {57}},\
  \bibinfo {pages} {1807} (\bibinfo {year} {2009})}\BibitemShut {NoStop}%
\bibitem [{\citenamefont {Shukla}\ \emph {et~al.}(2003)\citenamefont {Shukla},
  \citenamefont {Dwivedi},\ and\ \citenamefont {Stenflo}}]{Shukla_2003}%
  \BibitemOpen
  \bibfield  {author} {\bibinfo {author} {\bibfnamefont {P.}~\bibnamefont
  {Shukla}}, \bibinfo {author} {\bibfnamefont {P.}~\bibnamefont {Dwivedi}}, \
  and\ \bibinfo {author} {\bibfnamefont {L.}~\bibnamefont {Stenflo}},\
  }\href@noop {} {\bibfield  {journal} {\bibinfo  {journal} {New Journal of
  Physics}\ }\textbf {\bibinfo {volume} {5}},\ \bibinfo {pages} {22} (\bibinfo
  {year} {2003})}\BibitemShut {NoStop}%
\bibitem [{\citenamefont {Nekrasov}(2009)}]{nekrasov2009}%
  \BibitemOpen
  \bibfield  {author} {\bibinfo {author} {\bibfnamefont {A.}~\bibnamefont
  {Nekrasov}},\ }\href@noop {} {\bibfield  {journal} {\bibinfo  {journal} {The
  Astrophysical Journal}\ }\textbf {\bibinfo {volume} {695}},\ \bibinfo {pages}
  {46} (\bibinfo {year} {2009})}\BibitemShut {NoStop}%
\bibitem [{\citenamefont {Chabrier}\ \emph {et~al.}(2002)\citenamefont
  {Chabrier}, \citenamefont {Douchin},\ and\ \citenamefont
  {Potekhin}}]{chabrier2002dense}%
  \BibitemOpen
  \bibfield  {author} {\bibinfo {author} {\bibfnamefont {G.}~\bibnamefont
  {Chabrier}}, \bibinfo {author} {\bibfnamefont {F.}~\bibnamefont {Douchin}}, \
  and\ \bibinfo {author} {\bibfnamefont {A.}~\bibnamefont {Potekhin}},\
  }\href@noop {} {\bibfield  {journal} {\bibinfo  {journal} {Journal of
  Physics: Condensed Matter}\ }\textbf {\bibinfo {volume} {14}},\ \bibinfo
  {pages} {9133} (\bibinfo {year} {2002})}\BibitemShut {NoStop}%
\bibitem [{\citenamefont {Koester}\ and\ \citenamefont
  {Sch{\"o}nberner}(1986)}]{koester1986evolution}%
  \BibitemOpen
  \bibfield  {author} {\bibinfo {author} {\bibfnamefont {D.}~\bibnamefont
  {Koester}}\ and\ \bibinfo {author} {\bibfnamefont {D.}~\bibnamefont
  {Sch{\"o}nberner}},\ }\href@noop {} {\bibfield  {journal} {\bibinfo
  {journal} {Astronomy and Astrophysics}\ }\textbf {\bibinfo {volume} {154}},\
  \bibinfo {pages} {125} (\bibinfo {year} {1986})}\BibitemShut {NoStop}%
\bibitem [{\citenamefont {Willitsch}\ \emph {et~al.}(2008)\citenamefont
  {Willitsch}, \citenamefont {Bell}, \citenamefont {Gingell},\ and\
  \citenamefont {Softley}}]{willitsch2008chemical}%
  \BibitemOpen
  \bibfield  {author} {\bibinfo {author} {\bibfnamefont {S.}~\bibnamefont
  {Willitsch}}, \bibinfo {author} {\bibfnamefont {M.~T.}\ \bibnamefont {Bell}},
  \bibinfo {author} {\bibfnamefont {A.~D.}\ \bibnamefont {Gingell}}, \ and\
  \bibinfo {author} {\bibfnamefont {T.~P.}\ \bibnamefont {Softley}},\
  }\href@noop {} {\bibfield  {journal} {\bibinfo  {journal} {Physical Chemistry
  Chemical Physics}\ }\textbf {\bibinfo {volume} {10}},\ \bibinfo {pages}
  {7200} (\bibinfo {year} {2008})}\BibitemShut {NoStop}%
\bibitem [{\citenamefont {Pramanik}\ \emph {et~al.}(2002)\citenamefont
  {Pramanik}, \citenamefont {Prasad}, \citenamefont {Sen},\ and\ \citenamefont
  {Kaw}}]{pramanik2002experimental}%
  \BibitemOpen
  \bibfield  {author} {\bibinfo {author} {\bibfnamefont {J.}~\bibnamefont
  {Pramanik}}, \bibinfo {author} {\bibfnamefont {G.}~\bibnamefont {Prasad}},
  \bibinfo {author} {\bibfnamefont {A.}~\bibnamefont {Sen}}, \ and\ \bibinfo
  {author} {\bibfnamefont {P.}~\bibnamefont {Kaw}},\ }\href@noop {} {\bibfield
  {journal} {\bibinfo  {journal} {Physical review letters}\ }\textbf {\bibinfo
  {volume} {88}},\ \bibinfo {pages} {175001} (\bibinfo {year}
  {2002})}\BibitemShut {NoStop}%
\bibitem [{\citenamefont {Quinn}\ and\ \citenamefont
  {Goree}(2000)}]{quinn2000experimental}%
  \BibitemOpen
  \bibfield  {author} {\bibinfo {author} {\bibfnamefont {R.}~\bibnamefont
  {Quinn}}\ and\ \bibinfo {author} {\bibfnamefont {J.}~\bibnamefont {Goree}},\
  }\href@noop {} {\bibfield  {journal} {\bibinfo  {journal} {Physics of
  Plasmas}\ }\textbf {\bibinfo {volume} {7}},\ \bibinfo {pages} {3904}
  (\bibinfo {year} {2000})}\BibitemShut {NoStop}%
\bibitem [{\citenamefont {Ghosh}(2007)}]{ghosh2007rotation}%
  \BibitemOpen
  \bibfield  {author} {\bibinfo {author} {\bibfnamefont {P.}~\bibnamefont
  {Ghosh}},\ }\href@noop {} {\emph {\bibinfo {title} {Rotation and accretion
  powered pulsars}}},\ Vol.~\bibinfo {volume} {7}\ (\bibinfo  {publisher}
  {World Scientific},\ \bibinfo {year} {2007})\BibitemShut {NoStop}%
\bibitem [{\citenamefont {Hewish}\ \emph {et~al.}(1968)\citenamefont {Hewish},
  \citenamefont {Bell}, \citenamefont {Pilkington}, \citenamefont {Scott},\
  and\ \citenamefont {Collins}}]{hewish1968observation}%
  \BibitemOpen
  \bibfield  {author} {\bibinfo {author} {\bibfnamefont {A.}~\bibnamefont
  {Hewish}}, \bibinfo {author} {\bibfnamefont {S.~J.}\ \bibnamefont {Bell}},
  \bibinfo {author} {\bibfnamefont {J.~D.}\ \bibnamefont {Pilkington}},
  \bibinfo {author} {\bibfnamefont {P.~F.}\ \bibnamefont {Scott}}, \ and\
  \bibinfo {author} {\bibfnamefont {R.~A.}\ \bibnamefont {Collins}},\ }in\
  \href@noop {} {\emph {\bibinfo {booktitle} {Pulsating Stars}}}\ (\bibinfo
  {publisher} {Springer},\ \bibinfo {year} {1968})\ pp.\ \bibinfo {pages}
  {5--9}\BibitemShut {NoStop}%
\bibitem [{\citenamefont {Uchida}\ \emph {et~al.}(2004)\citenamefont {Uchida},
  \citenamefont {Konopka},\ and\ \citenamefont {Morfill}}]{uchida2004}%
  \BibitemOpen
  \bibfield  {author} {\bibinfo {author} {\bibfnamefont {G.}~\bibnamefont
  {Uchida}}, \bibinfo {author} {\bibfnamefont {U.}~\bibnamefont {Konopka}}, \
  and\ \bibinfo {author} {\bibfnamefont {G.}~\bibnamefont {Morfill}},\
  }\href@noop {} {\bibfield  {journal} {\bibinfo  {journal} {Physical review
  letters}\ }\textbf {\bibinfo {volume} {93}},\ \bibinfo {pages} {155002}
  (\bibinfo {year} {2004})}\BibitemShut {NoStop}%
\bibitem [{\citenamefont {Thomas}\ \emph {et~al.}(2015)\citenamefont {Thomas},
  \citenamefont {Konopka}, \citenamefont {Artis}, \citenamefont {Lynch},
  \citenamefont {Leblanc}, \citenamefont {Adams}, \citenamefont {Merlino},\
  and\ \citenamefont {Rosenberg}}]{thomas2015magnetized}%
  \BibitemOpen
  \bibfield  {author} {\bibinfo {author} {\bibfnamefont {E.}~\bibnamefont
  {Thomas}}, \bibinfo {author} {\bibfnamefont {U.}~\bibnamefont {Konopka}},
  \bibinfo {author} {\bibfnamefont {D.}~\bibnamefont {Artis}}, \bibinfo
  {author} {\bibfnamefont {B.}~\bibnamefont {Lynch}}, \bibinfo {author}
  {\bibfnamefont {S.}~\bibnamefont {Leblanc}}, \bibinfo {author} {\bibfnamefont
  {S.}~\bibnamefont {Adams}}, \bibinfo {author} {\bibfnamefont
  {R.}~\bibnamefont {Merlino}}, \ and\ \bibinfo {author} {\bibfnamefont
  {M.}~\bibnamefont {Rosenberg}},\ }\href@noop {} {\bibfield  {journal}
  {\bibinfo  {journal} {Journal of Plasma Physics}\ }\textbf {\bibinfo {volume}
  {81}} (\bibinfo {year} {2015})}\BibitemShut {NoStop}%
\bibitem [{\citenamefont {Thomas}\ \emph {et~al.}(2012)\citenamefont {Thomas},
  \citenamefont {Merlino},\ and\ \citenamefont
  {Rosenberg}}]{thomas2012magnetized}%
  \BibitemOpen
  \bibfield  {author} {\bibinfo {author} {\bibfnamefont {E.}~\bibnamefont
  {Thomas}}, \bibinfo {author} {\bibfnamefont {R.}~\bibnamefont {Merlino}}, \
  and\ \bibinfo {author} {\bibfnamefont {M.}~\bibnamefont {Rosenberg}},\
  }\href@noop {} {\bibfield  {journal} {\bibinfo  {journal} {Plasma Physics and
  Controlled Fusion}\ }\textbf {\bibinfo {volume} {54}},\ \bibinfo {pages}
  {124034} (\bibinfo {year} {2012})}\BibitemShut {NoStop}%
\bibitem [{\citenamefont {K{\"a}hlert}\ \emph {et~al.}(2012)\citenamefont
  {K{\"a}hlert}, \citenamefont {Carstensen}, \citenamefont {Bonitz},
  \citenamefont {L{\"o}wen}, \citenamefont {Greiner},\ and\ \citenamefont
  {Piel}}]{kahlert2012}%
  \BibitemOpen
  \bibfield  {author} {\bibinfo {author} {\bibfnamefont {H.}~\bibnamefont
  {K{\"a}hlert}}, \bibinfo {author} {\bibfnamefont {J.}~\bibnamefont
  {Carstensen}}, \bibinfo {author} {\bibfnamefont {M.}~\bibnamefont {Bonitz}},
  \bibinfo {author} {\bibfnamefont {H.}~\bibnamefont {L{\"o}wen}}, \bibinfo
  {author} {\bibfnamefont {F.}~\bibnamefont {Greiner}}, \ and\ \bibinfo
  {author} {\bibfnamefont {A.}~\bibnamefont {Piel}},\ }\href@noop {} {\bibfield
   {journal} {\bibinfo  {journal} {Physical review letters}\ }\textbf {\bibinfo
  {volume} {109}},\ \bibinfo {pages} {155003} (\bibinfo {year}
  {2012})}\BibitemShut {NoStop}%
\bibitem [{\citenamefont {Hartmann}\ \emph {et~al.}(2013)\citenamefont
  {Hartmann}, \citenamefont {Donk{\'o}}, \citenamefont {Ott}, \citenamefont
  {K{\"a}hlert},\ and\ \citenamefont {Bonitz}}]{hartmann2013}%
  \BibitemOpen
  \bibfield  {author} {\bibinfo {author} {\bibfnamefont {P.}~\bibnamefont
  {Hartmann}}, \bibinfo {author} {\bibfnamefont {Z.}~\bibnamefont {Donk{\'o}}},
  \bibinfo {author} {\bibfnamefont {T.}~\bibnamefont {Ott}}, \bibinfo {author}
  {\bibfnamefont {H.}~\bibnamefont {K{\"a}hlert}}, \ and\ \bibinfo {author}
  {\bibfnamefont {M.}~\bibnamefont {Bonitz}},\ }\href@noop {} {\bibfield
  {journal} {\bibinfo  {journal} {Physical Review Letters}\ }\textbf {\bibinfo
  {volume} {111}},\ \bibinfo {pages} {155002} (\bibinfo {year}
  {2013})}\BibitemShut {NoStop}%
\bibitem [{\citenamefont {kaur}\ \emph {et~al.}(2015)\citenamefont {kaur},
  \citenamefont {Bose}, \citenamefont {Chattopadhyay}, \citenamefont {Sharma},
  \citenamefont {Ghosh}, \citenamefont {Saxena},\ and\ \citenamefont
  {Thomas}}]{manjit}%
  \BibitemOpen
  \bibfield  {author} {\bibinfo {author} {\bibfnamefont {M.}~\bibnamefont
  {kaur}}, \bibinfo {author} {\bibfnamefont {S.}~\bibnamefont {Bose}}, \bibinfo
  {author} {\bibfnamefont {P.}~\bibnamefont {Chattopadhyay}}, \bibinfo {author}
  {\bibfnamefont {D.}~\bibnamefont {Sharma}}, \bibinfo {author} {\bibfnamefont
  {J.}~\bibnamefont {Ghosh}}, \bibinfo {author} {\bibfnamefont {Y.~C.}\
  \bibnamefont {Saxena}}, \ and\ \bibinfo {author} {\bibfnamefont {E.~J.}\
  \bibnamefont {Thomas}},\ }\href@noop {} {\bibfield  {journal} {\bibinfo
  {journal} {Physics of Plasmas}\ }\textbf {\bibinfo {volume} {22}},\ \bibinfo
  {pages} {093702} (\bibinfo {year} {2015})}\BibitemShut {NoStop}%
\bibitem [{\citenamefont {Kumar}\ and\ \citenamefont {Sharma}(2020)}]{prince}%
  \BibitemOpen
  \bibfield  {author} {\bibinfo {author} {\bibfnamefont {P.}~\bibnamefont
  {Kumar}}\ and\ \bibinfo {author} {\bibfnamefont {D.}~\bibnamefont {Sharma}},\
  }\href@noop {} {\bibfield  {journal} {\bibinfo  {journal} {Physics of
  Plasmas}\ }\textbf {\bibinfo {volume} {27}},\ \bibinfo {pages} {063703}
  (\bibinfo {year} {2020})}\BibitemShut {NoStop}%
\bibitem [{\citenamefont {Golden}\ and\ \citenamefont {Kalman}(2000)}]{Golden}%
  \BibitemOpen
  \bibfield  {author} {\bibinfo {author} {\bibfnamefont {K.~I.}\ \bibnamefont
  {Golden}}\ and\ \bibinfo {author} {\bibfnamefont {G.~J.}\ \bibnamefont
  {Kalman}},\ }\href@noop {} {\bibfield  {journal} {\bibinfo  {journal}
  {Physics of Plasmas}\ }\textbf {\bibinfo {volume} {7}},\ \bibinfo {pages}
  {14} (\bibinfo {year} {2000})}\BibitemShut {NoStop}%
\bibitem [{\citenamefont {Kalman}\ and\ \citenamefont
  {Golden}(1990)}]{kalman1990response}%
  \BibitemOpen
  \bibfield  {author} {\bibinfo {author} {\bibfnamefont {G.}~\bibnamefont
  {Kalman}}\ and\ \bibinfo {author} {\bibfnamefont {K.}~\bibnamefont
  {Golden}},\ }\href@noop {} {\bibfield  {journal} {\bibinfo  {journal}
  {Physical Review A}\ }\textbf {\bibinfo {volume} {41}},\ \bibinfo {pages}
  {5516} (\bibinfo {year} {1990})}\BibitemShut {NoStop}%
\bibitem [{\citenamefont {Golden}\ \emph {et~al.}(1992)\citenamefont {Golden},
  \citenamefont {Kalman},\ and\ \citenamefont {Wyns}}]{golden1992dielectric}%
  \BibitemOpen
  \bibfield  {author} {\bibinfo {author} {\bibfnamefont {K.~I.}\ \bibnamefont
  {Golden}}, \bibinfo {author} {\bibfnamefont {G.}~\bibnamefont {Kalman}}, \
  and\ \bibinfo {author} {\bibfnamefont {P.}~\bibnamefont {Wyns}},\ }\href@noop
  {} {\bibfield  {journal} {\bibinfo  {journal} {Physical Review A}\ }\textbf
  {\bibinfo {volume} {46}},\ \bibinfo {pages} {3454} (\bibinfo {year}
  {1992})}\BibitemShut {NoStop}%
\bibitem [{\citenamefont {Hou}\ \emph {et~al.}(2009)\citenamefont {Hou},
  \citenamefont {Mi{\v{s}}kovi{\'c}}, \citenamefont {Piel},\ and\ \citenamefont
  {Murillo}}]{hou2009wave}%
  \BibitemOpen
  \bibfield  {author} {\bibinfo {author} {\bibfnamefont {L.-J.}\ \bibnamefont
  {Hou}}, \bibinfo {author} {\bibfnamefont {Z.}~\bibnamefont
  {Mi{\v{s}}kovi{\'c}}}, \bibinfo {author} {\bibfnamefont {A.}~\bibnamefont
  {Piel}}, \ and\ \bibinfo {author} {\bibfnamefont {M.~S.}\ \bibnamefont
  {Murillo}},\ }\href@noop {} {\bibfield  {journal} {\bibinfo  {journal}
  {Physical Review E}\ }\textbf {\bibinfo {volume} {79}},\ \bibinfo {pages}
  {046412} (\bibinfo {year} {2009})}\BibitemShut {NoStop}%
\bibitem [{\citenamefont {Jiang}\ \emph {et~al.}(2007)\citenamefont {Jiang},
  \citenamefont {Song},\ and\ \citenamefont {Wang}}]{jiang2007theoretical}%
  \BibitemOpen
  \bibfield  {author} {\bibinfo {author} {\bibfnamefont {K.}~\bibnamefont
  {Jiang}}, \bibinfo {author} {\bibfnamefont {Y.-H.}\ \bibnamefont {Song}}, \
  and\ \bibinfo {author} {\bibfnamefont {Y.-N.}\ \bibnamefont {Wang}},\
  }\href@noop {} {\bibfield  {journal} {\bibinfo  {journal} {Physics of
  Plasmas}\ }\textbf {\bibinfo {volume} {14}},\ \bibinfo {pages} {103708}
  (\bibinfo {year} {2007})}\BibitemShut {NoStop}%
\bibitem [{\citenamefont {Khrapak}\ \emph
  {et~al.}(2016{\natexlab{a}})\citenamefont {Khrapak}, \citenamefont
  {Molotkov}, \citenamefont {Lipaev}, \citenamefont {Zhukhovitskii},
  \citenamefont {Naumkin}, \citenamefont {Fortov}, \citenamefont {Petrov},
  \citenamefont {Thomas}, \citenamefont {Khrapak}, \citenamefont {Huber} \emph
  {et~al.}}]{khrapak2016complex}%
  \BibitemOpen
  \bibfield  {author} {\bibinfo {author} {\bibfnamefont {A.}~\bibnamefont
  {Khrapak}}, \bibinfo {author} {\bibfnamefont {V.}~\bibnamefont {Molotkov}},
  \bibinfo {author} {\bibfnamefont {A.}~\bibnamefont {Lipaev}}, \bibinfo
  {author} {\bibfnamefont {D.}~\bibnamefont {Zhukhovitskii}}, \bibinfo {author}
  {\bibfnamefont {V.}~\bibnamefont {Naumkin}}, \bibinfo {author} {\bibfnamefont
  {V.}~\bibnamefont {Fortov}}, \bibinfo {author} {\bibfnamefont
  {O.}~\bibnamefont {Petrov}}, \bibinfo {author} {\bibfnamefont
  {H.}~\bibnamefont {Thomas}}, \bibinfo {author} {\bibfnamefont
  {S.}~\bibnamefont {Khrapak}}, \bibinfo {author} {\bibfnamefont
  {P.}~\bibnamefont {Huber}},  \emph {et~al.},\ }\href@noop {} {\enquote
  {\bibinfo {title} {Complex plasma research under microgravity conditions:
  Pk-3 plus laboratory on the international space station},}\ } (\bibinfo
  {year} {2016}{\natexlab{a}})\BibitemShut {NoStop}%
\bibitem [{\citenamefont {Pustylnik}\ \emph {et~al.}(2016)\citenamefont
  {Pustylnik}, \citenamefont {Fink}, \citenamefont {Nosenko}, \citenamefont
  {Antonova}, \citenamefont {Hagl}, \citenamefont {Thomas}, \citenamefont
  {Zobnin}, \citenamefont {Lipaev}, \citenamefont {Usachev}, \citenamefont
  {Molotkov} \emph {et~al.}}]{pustylnik2016plasmakristall}%
  \BibitemOpen
  \bibfield  {author} {\bibinfo {author} {\bibfnamefont {M.}~\bibnamefont
  {Pustylnik}}, \bibinfo {author} {\bibfnamefont {M.}~\bibnamefont {Fink}},
  \bibinfo {author} {\bibfnamefont {V.}~\bibnamefont {Nosenko}}, \bibinfo
  {author} {\bibfnamefont {T.}~\bibnamefont {Antonova}}, \bibinfo {author}
  {\bibfnamefont {T.}~\bibnamefont {Hagl}}, \bibinfo {author} {\bibfnamefont
  {H.}~\bibnamefont {Thomas}}, \bibinfo {author} {\bibfnamefont
  {A.}~\bibnamefont {Zobnin}}, \bibinfo {author} {\bibfnamefont
  {A.}~\bibnamefont {Lipaev}}, \bibinfo {author} {\bibfnamefont
  {A.}~\bibnamefont {Usachev}}, \bibinfo {author} {\bibfnamefont
  {V.}~\bibnamefont {Molotkov}},  \emph {et~al.},\ }\href@noop {} {\bibfield
  {journal} {\bibinfo  {journal} {Review of scientific instruments}\ }\textbf
  {\bibinfo {volume} {87}},\ \bibinfo {pages} {093505} (\bibinfo {year}
  {2016})}\BibitemShut {NoStop}%
\bibitem [{\citenamefont {Fortov}\ \emph {et~al.}(2003)\citenamefont {Fortov},
  \citenamefont {Vaulina}, \citenamefont {Petrov}, \citenamefont {Molotkov},
  \citenamefont {Chernyshev}, \citenamefont {Lipaev}, \citenamefont {Morfill},
  \citenamefont {Thomas}, \citenamefont {Rotermell}, \citenamefont {Khrapak}
  \emph {et~al.}}]{fortov2003dynamics}%
  \BibitemOpen
  \bibfield  {author} {\bibinfo {author} {\bibfnamefont {V.}~\bibnamefont
  {Fortov}}, \bibinfo {author} {\bibfnamefont {O.}~\bibnamefont {Vaulina}},
  \bibinfo {author} {\bibfnamefont {O.}~\bibnamefont {Petrov}}, \bibinfo
  {author} {\bibfnamefont {V.}~\bibnamefont {Molotkov}}, \bibinfo {author}
  {\bibfnamefont {A.}~\bibnamefont {Chernyshev}}, \bibinfo {author}
  {\bibfnamefont {A.}~\bibnamefont {Lipaev}}, \bibinfo {author} {\bibfnamefont
  {G.}~\bibnamefont {Morfill}}, \bibinfo {author} {\bibfnamefont
  {H.}~\bibnamefont {Thomas}}, \bibinfo {author} {\bibfnamefont
  {H.}~\bibnamefont {Rotermell}}, \bibinfo {author} {\bibfnamefont
  {S.}~\bibnamefont {Khrapak}},  \emph {et~al.},\ }\href@noop {} {\bibfield
  {journal} {\bibinfo  {journal} {Journal of Experimental and Theoretical
  Physics}\ }\textbf {\bibinfo {volume} {96}},\ \bibinfo {pages} {704}
  (\bibinfo {year} {2003})}\BibitemShut {NoStop}%
\bibitem [{\citenamefont {Golden}\ \emph {et~al.}(1993)\citenamefont {Golden},
  \citenamefont {Kalman},\ and\ \citenamefont {Wyns}}]{golden1993dielectric}%
  \BibitemOpen
  \bibfield  {author} {\bibinfo {author} {\bibfnamefont {K.~I.}\ \bibnamefont
  {Golden}}, \bibinfo {author} {\bibfnamefont {G.}~\bibnamefont {Kalman}}, \
  and\ \bibinfo {author} {\bibfnamefont {P.}~\bibnamefont {Wyns}},\ }\href@noop
  {} {\bibfield  {journal} {\bibinfo  {journal} {Physical Review B}\ }\textbf
  {\bibinfo {volume} {48}},\ \bibinfo {pages} {8882} (\bibinfo {year}
  {1993})}\BibitemShut {NoStop}%
\bibitem [{\citenamefont {Khrapak}\ \emph
  {et~al.}(2016{\natexlab{b}})\citenamefont {Khrapak}, \citenamefont {Klumov},
  \citenamefont {Couedel},\ and\ \citenamefont {Thomas}}]{khrapak2016long}%
  \BibitemOpen
  \bibfield  {author} {\bibinfo {author} {\bibfnamefont {S.~A.}\ \bibnamefont
  {Khrapak}}, \bibinfo {author} {\bibfnamefont {B.}~\bibnamefont {Klumov}},
  \bibinfo {author} {\bibfnamefont {L.}~\bibnamefont {Couedel}}, \ and\
  \bibinfo {author} {\bibfnamefont {H.~M.}\ \bibnamefont {Thomas}},\
  }\href@noop {} {\bibfield  {journal} {\bibinfo  {journal} {Physics of
  Plasmas}\ }\textbf {\bibinfo {volume} {23}},\ \bibinfo {pages} {023702}
  (\bibinfo {year} {2016}{\natexlab{b}})}\BibitemShut {NoStop}%
\bibitem [{\citenamefont {Khrapak}(2017)}]{khrapak2017practical}%
  \BibitemOpen
  \bibfield  {author} {\bibinfo {author} {\bibfnamefont {S.~A.}\ \bibnamefont
  {Khrapak}},\ }\href@noop {} {\bibfield  {journal} {\bibinfo  {journal} {AIP
  Advances}\ }\textbf {\bibinfo {volume} {7}},\ \bibinfo {pages} {125026}
  (\bibinfo {year} {2017})}\BibitemShut {NoStop}%
\bibitem [{\citenamefont {Khrapak}\ \emph {et~al.}(2019)\citenamefont
  {Khrapak}, \citenamefont {Khrapak}, \citenamefont {Kryuchkov},\ and\
  \citenamefont {Yurchenko}}]{khrapak2019onset}%
  \BibitemOpen
  \bibfield  {author} {\bibinfo {author} {\bibfnamefont {S.~A.}\ \bibnamefont
  {Khrapak}}, \bibinfo {author} {\bibfnamefont {A.~G.}\ \bibnamefont
  {Khrapak}}, \bibinfo {author} {\bibfnamefont {N.~P.}\ \bibnamefont
  {Kryuchkov}}, \ and\ \bibinfo {author} {\bibfnamefont {S.~O.}\ \bibnamefont
  {Yurchenko}},\ }\href@noop {} {\bibfield  {journal} {\bibinfo  {journal} {The
  Journal of chemical physics}\ }\textbf {\bibinfo {volume} {150}},\ \bibinfo
  {pages} {104503} (\bibinfo {year} {2019})}\BibitemShut {NoStop}%
\end{thebibliography}%


\end{document}